\documentclass[]{article}
\usepackage{graphicx}
\usepackage{amsmath}
\usepackage{url}
\usepackage{cite}
\usepackage{authblk}
\usepackage[font=small]{caption}
\usepackage{hyperref}
\usepackage{lmodern}
\usepackage{amssymb,amsmath}
\usepackage{ifxetex,ifluatex}
\usepackage{fixltx2e} 
\ifnum 0\ifxetex 1\fi\ifluatex 1\fi=0 
  \usepackage[T1]{fontenc}
  \usepackage[utf8]{inputenc}
\else 
  \ifxetex
    \usepackage{mathspec}
  \else
    \usepackage{fontspec}
  \fi
  \defaultfontfeatures{Ligatures=TeX,Scale=MatchLowercase}
\fi
\IfFileExists{upquote.sty}{\usepackage{upquote}}{}
\IfFileExists{microtype.sty}{%
\usepackage{microtype}
\UseMicrotypeSet[protrusion]{basicmath} 
}{}
\usepackage[margin=1in]{geometry}
\usepackage{hyperref}
\hypersetup{unicode=true,
            pdftitle={How different homophily preferences mitigate and spur ethnic and value segregation: Schelling's model extended},
            pdfauthor={Rocco Paolillo, Jan Lorenz},
            pdfborder={0 0 0},
            breaklinks=true}
\usepackage{longtable,booktabs}
\usepackage{graphicx,grffile}
\makeatletter
\def\maxwidth{\ifdim\Gin@nat@width>\linewidth\linewidth\else\Gin@nat@width\fi}
\def\maxheight{\ifdim\Gin@nat@height>\textheight\textheight\else\Gin@nat@height\fi}
\makeatother
\setkeys{Gin}{width=\maxwidth,height=\maxheight,keepaspectratio}
\IfFileExists{parskip.sty}{%
\usepackage{parskip}
}{
\setlength{\parindent}{0pt}
\setlength{\parskip}{6pt plus 2pt minus 1pt}
}
\ifx\paragraph\undefined\else
\let\oldparagraph\paragraph
\renewcommand{\paragraph}[1]{\oldparagraph{#1}\mbox{}}
\fi
\ifx\subparagraph\undefined\else
\let\oldsubparagraph\subparagraph
\renewcommand{\subparagraph}[1]{\oldsubparagraph{#1}\mbox{}}
\fi

\let\rmarkdownfootnote\footnote%
\def\footnote{\protect\rmarkdownfootnote}

\usepackage{titling}

  \title{How different homophily preferences mitigate and spur ethnic and value
segregation: Schelling's model extended}
  \pretitle{\vspace{\droptitle}\centering\huge}
  \posttitle{\par}
  \author{Rocco Paolillo$^{1,\dagger}$ \& Jan Lorenz$^{1,2\ddagger}$}
\affil{$^{1}$Bremen International Graduate School of Social Sciences, Jacobs University Bremen}
\affil{$^{2}$Computational Social Science, GESIS Leibniz Institute for the Social Sciences}
\affil{$^{\dagger}$rpaolillo@bigsss-bremen.de}
\affil{$^{\ddagger}$j.lorenz@jacobs-university.de}
 \preauthor{\centering\large\emph}
  \postauthor{\par}

\usepackage{float}
\usepackage{multirow}

\usepackage{amsthm}

\theoremstyle{definition}

\theoremstyle{definition}

\theoremstyle{definition}

\theoremstyle{remark}

\date{}
\begin{document}
\maketitle

\begin{abstract}
In Schelling's segregation model agents of two ethnic groups reside in a regular grid and aim to live in a neighborhood that matches the minimum desired fraction of members of the same ethnicity. The model shows that observed segregation can be the emergent result of people interacting under spatial constraints in pursuit of such homophily preferences. Even mild homophily preferences can generate high degrees of segregation at the macro level. In modern, ethnically diverse societies people might not define similarity based on ethnicity. Instead, shared tolerance towards ethnic diversity might play a more significant role, impacting segregation and integration patterns in societies. Bearing this consideration in mind, we extend Schelling's model by dividing the population of agents into value-oriented and ethnicity-oriented agents. Using parameter sweeping, we explore the consequences that the mutual adaptation of these two types of agents has on ethnic segregation, value segregation, and population density in the neighborhood. Such consequences are examined for equally sized ethnic groups and for majority-minority conditions. The introduction of value-oriented agents reduces total ethnic segregation when compared to Schelling's original model, but the new phenomenon of value segregation appears to be more pronounced than ethnic segregation. Furthermore, due to cross-contagion, stronger ethnic homophily preferences lead not only to greater ethnic segregation but also to more value segregation. Stronger value-orientation of the tolerant agents similarly leads to increased ethnic segregation of the ethnicity-oriented agents. Also, value-oriented agents tend to live in neighborhoods with more agents than ethnicity-oriented agents. In majority-minority settings, such effects appear to be more drastic for the minority than the majority ethnicity.\\


\end{abstract}


\section{Introduction}
Residential segregation can be defined as the spatial and unequal distribution of different groups in a city \cite{clark2015}. This topic is critical for policymakers and social research because it reduces the probability of interaction between diverse groups \cite{massey88}. Regarding social integration, the reduced likelihood of interaction is considered to be a major obstacle to minority participation in the mainstream society (e.g. immigrants) \cite{Esser2010}. Therefore, acceptance of diversity is an essential ingredient of the multidimensional concept of social cohesion \cite{DragolovIgnaczea2016SocialCohesionin}. Although there is no unanimous agreement that residential proximity is associated with successful integration, researchers are mainly interested in the reasons for this observed segregation \cite{bolt2010}. Two main explanations are that first, the physical distance between diverse groups is a reflection of their social distance, and second, this distance is the result of local residents' active strategies to oppose ethnic minorities \cite{clark2008, bruch2014}. On the contrary, the theoretical observation of Thomas Schelling \cite{schelling69} is that segregation might  be the emergent result of the mutual adaptation of people who seek to live close to members of their own group within spatial constraints, without implying negative attitudes towards other groups. Although the term was not used by Schelling, we can define these individual choices as homophily preferences. Homophily is the tendency to associate with those that are considered similar \cite{mcpherson2001}. Even though this concept is mainly used in the field of network science, it can be applied to aggregation in neighborhoods. As a matter of fact, some studies on segregation dynamics based on Schelling's work expressely consider homophily as a driver of residential moves and neighborhood preferences \cite{muller2018, magi2016}. Schelling's model is built on the definition of similarity as a "twofold, exhaustive and recognizable distinction" \cite[p.~488]{schelling69}, which he applied to the racial segregation of black and white ethnic groups in the U.S. census. The population in the model is divided into two groups, whose members live in a neighborhood as long as its composition matches the desired fraction of perceived similar people \cite{schelling69, schelling71}. The main contribution of Schelling's work is that it shows how segregation at the societal level can emerge as a stable outcome from the "interplay of individual choices" \cite[p.~488]{schelling69}, even from less desired fractions \cite{schelling69, schelling71}. Nowadays, Schelling's model is recognized as a seminal agent-based model \cite{hatna2015} and one of the first contributions in computational social sciences and complex systems dynamics \cite{epstein1996}. In the field of residential segregation, it is often cited as a prototype of checkerboard models \cite{zhang2011} and tipping behavior in residential dynamics \cite{card2008}.\\
In this paper, we focus on the concept of similarity in Schelling's model. The definition of similarity is based on a "twofold, exhaustive and recognizable distinction" \cite[p.~488]{schelling69} and has  been associated with a binary definition of ethnic segregation \cite{clark2015, zhang2011}. Nevertheless, as already Schelling pointed out, similarity does not need to be based on ethnicity \cite{schelling69, schelling71}. In modern societies, people might be part of diverse social groups \cite{roccas2002}. In a multicultural context, this diversity might include not only different ethnic backgrounds, but also different values \cite{schwartz2012} or attitudes towards integration \cite{bourhis1997}. In these societies, people might differ in their way of constructing similarity out of the different categories that they are a part of. According to the ethnic boundary making perspective, in multicultural contexts, distinctions between different groups derive from the reciprocal construction of symbolic boundaries distinguishing "us" and "them" \cite{bail2008}. This process emerges from the constructivist, subjective and interactive definitions of similarity and inclusion that people engage in, independent of their ethnic membership \cite{wimmer2007, wimmer13}. Thus, homophily preferences assoociated with these dynamics can be conceptualized in various ways, which can have interesting implications for segregation scenarios. Lazarsfeld and Merton \cite{lazarsfeld1954}, as reviewed by McPherson et al. \cite{mcpherson2001}, distinguished between two determinants of homophily behavior. Status homophily is based on characteristics that stratify society, such as ethnicity, whereas value homophily is grounded in internal status such as attitudes and opinions \cite{mcpherson2001}. We focus here on homophily that is created through shared values of tolerance for two reasons. On the one hand, it has a direct effect on the acceptance of an ethnically diverse neighborhood \cite{marjoka2014} and favors contacts between groups \cite{brewer2005}. On the other hand, tolerance can cause conflicts with members of an individual's group who do not share similar attitudes, so that the mismatch of attitudes can threaten the group's unity \cite{verkuyten2010}. When tolerant agents subscribe to Karl Popper \cite{popper45}, they should not tolerate intolerance, which is called the "Paradox of Tolerance." Similar to the agents in Schelling's original model, these tolerant agents would prefer to live in a neighborhood with at least some others who share their tolerance values.\\
We are additionally interested in the role of relative group sizes, which did not receive much attention in Schelling's demonstrations \cite{schelling71}. Indeed, most of the following studies typically assume equal sizes of ethnic groups in the model \cite{troitzsch2017}. However, the equal distribution of different groups is likely to not be the case in multi-ethnic societies. As a matter of fact, relative group sizes are an important topic in segregation studies, because they change the likelihood of interaction between members of the own and other groups. The main interest is in the consequences that such uneven probabilities can have on integration between groups \cite{massey88}. Regarding migrant integration, Esser \cite{Esser2010} suggests that the smaller the group, the greater the chance of interacting with and assimilating into the local population. Nevertheless, isolation can favor segregation over assimilation by reducing the actual chances of interaction \cite{massey88}. To the best of our knowledge, there is only one study by Troitzsch \cite{troitzsch2017} that simulates minority and majority group sizes in Schelling's model and the impact of desired fraction of each group on the overall emerged segregation. We therefore consider of additional interest to investigate the role of group sizes combined with different homophily preferences such as ethnic and value homophily.\\
In  sum, we are interested in the consequences that different homophily preferences, here ethnic homophily and value homophily, can have on emergent segregation in Schelling's model. To this aim, we extend  the model by dividing the population of both ethnic groups into value-oriented and ethnicity-oriented agents. Utilizing parameter sweeping, we determine what outcomes of ethnic segregation, value segregation, and population density in the neighborhood emerge from different homophily thresholds with agents in equal group sizes and majority-minority conditions.

\section{Model and Simulation Experiments}
Agents represent individuals who are relocating. Each agent is defined by two overlapping static state variables: {\it ethnicity}, as in Schelling's original model, and {\it value orientation}. Ethnicity is modeled by a color tag, whereby each agent is assigned to either ethnicity 1 (blue), or ethnicity 2 (orange). Value attribution is modeled through a shape tag, dividing agents into ethnicity-oriented agents (square) and value-oriented agents (circle). Agents, based on their value attribution, use different criteria to define their homophily preferences. Value-oriented agents are tolerant with respect to ethnicity but not tolerant to intolerance. This means that they consider agents are similar when they share the same tolerance values; they ignore ethnicity. Ethnicity-oriented agents are intolerant, meaning they consider agents are similar when they share their ethnicity; these agents ignore the level of tolerance. Homophily preferences do not change during the simulation.

We implemented our model by extending NetLogo's version of Schelling's model \cite{Wilenskysm}, specifically version NetLogo 6.0.4 \cite{Wilenskynl}, and keeping the standard periodic boundary conditions of the model (torus world) within a 51 times 51 patches regular grid. Thus, there are 2601 patches where agents can live, and each patch has eight neighboring patches (Moore neighborhood). On each patch, there is space for one agent to live. 

The initial configuration is based on two parameters: the expected {\it density} and the {\it fraction of the majority ethnicity}. Both ethnic groups are divided in half value-oriented and half ethnicity-oriented agents.
The initialization goes as follows. For each patch, an agent is positioned there with a probability determined by the expected density. The ethnicity (color) of the agent is set to blue, with probability determined by the fraction of the majority ethnicity. A final random draw, with a 0.5 probability, decides the value orientation (shape), which is set to either a square or a circle. For agent $i$, we call her ethnicity $E(i)$ and her value orientation $V(i)$. 

Global parameters for a simulation run are the {\it ethnic homophily threshold} and the {\it value homophily threshold} $\theta^\text{E}, \theta^\text{V} \in [0,1]$. For the ethnicity-oriented agents, the ethnic homophily threshold determines the minimum fraction of agents with the same ethnicity that they want in their neighborhoods. For the value-oriented agents, analog, the value homophily threshold determines the minimum fraction of other value-oriented agents that they want in their neighborhoods. We define 
\begin{equation}
\Theta^\text{E}_i = \frac{\#\{j \in N(i) | E(j)=E(i)\}}{\#N(i)}
\end{equation} 
as the {\it ethnic segregation in the neighborhood} of agent $i$ and \\
\begin{equation}
\Theta^\text{V}_i = \frac{\#\{j \in N(i) | V(j)=V(i)\}}{\#N(i)}
\end{equation}
as the {\it value segregation in the neighborhood}, where $\#N(i)$ stands for the number of agents $j$ in her current Moore neighborhood.\\ 

An ethnicity-oriented agent $i$ is {\it happy} when $\Theta^\text{E}_i \geq \theta^\text{E}$. A value-oriented agent is {\it happy} when $\Theta^\text{V}_i \geq \theta^\text{V}$. Agents are also happy when their neighborhood is empty. Fig. \ref{fig:AnalysisMapExamples}A shows the different types of agents in our model, demonstrating which other types of agents in the neighborhood contribute to the happiness of ethnicity-oriented and value-oriented agents.
The dynamic flow of the model is as follows.  
At each time step after initialization, all unhappy agents randomly relocate to a free spot. There is no targeted relocation. We kept the recursive relocation mechanism of the NetLogo implementation. Then, the happiness state of all agents is updated based on recalculation of $\Theta^\text{E}_i$ and $\Theta^\text{V}_i$.\\
\\

\begin{table}

\centering
\begin{tabular}{lll}
 Static agent variable                      & Range                               \\ 
\midrule \midrule
 Ethnicity $E(i)$  (\texttt{color})         & Ethnicity 1 (\texttt{blue}), Ethnicity 2 (\texttt{orange})         \\
 Value orientation $V(i)$  (\texttt{shape}) & ethnicity-oriented  (\texttt{square}), value-oriented  (\texttt{circle}) \\
 \bottomrule
\end{tabular}

\bigskip

\begin{tabular}{lll}
 Parameters for setup of initial conditions    & Range   & Parameter sweep  \\ \midrule\midrule
 Population density (\texttt{density})                       & 0.5--0.99   & 0.7          \\
 Fraction of majority ethnicity (\texttt{fraction\_majority}) & 0.5--1      & $0.5,\stackrel{+0.1}{\dots}, 0.9$  \\
 \bottomrule
\end{tabular}

\bigskip

\begin{tabular}{lll}
 Global parameters for simulation run    & Range   & Parameter sweep  \\ \midrule\midrule
 Ethnic homophily threshold $\theta^\text{E}$ (\texttt{ethnic\_homophily}) & [0,1]   & $0,\stackrel{+0.1}{\dots}, 1$ \\
 Value homophily threshold $\theta^\text{V}$ (\texttt{value\_homophily}) & [0,1]   & $0,\stackrel{+0.1}{\dots}, 1$  \\
 \bottomrule
\end{tabular}
\bigskip

\begin{tabular}{lll}
 Dynamic variables for agents                 & Computation & Range                \\ \midrule\midrule
Ethnic similarity in neighborhood  $\Theta^\text{E}_i$ &  $\#\{j \in N(i) | E(j)=E(i)\} / \#N(i)$ & [0,1] \\ 
 Value similarity in neighborhood $\Theta^\text{V}_i$ & $\#\{j \in N(i) | V(j)=V(i)\} / \#N(i)$ & [0,1]  \\ 
 \multirow{2}{*}{Happiness of $i$ (\texttt{happy?})} & $\Theta^\text{E}_i \geq \theta^\text{E}$ if $V(i)$ = ethnicity-oriented, & \multirow{2}{*}{TRUE, FALSE} \\
  & $\Theta^\text{V}_i \geq \theta^\text{V}$ if $V(i)$ = value-oriented &   \\
 Neighborhood density $d_i$           & $d_i = \frac{\#N(i)}{8}$ & [0,1]  \\
 \bottomrule
\end{tabular}

\bigskip

\begin{tabular}{lll}
 Global output measures    & Computation   \\ \midrule\midrule
 Fraction of happy agents & \# happy agents / \# agents \\
 Ethnic segregation $\Theta^\text{E}$       & mean ethnic segregation      \\ 
  ... of all agents          & (Figs. \ref{fig:OverallSegregation}, \ref{fig:AnalysisMapExamples},  and \ref{fig:Heatmap})  \\
  ... of ethnicity-oriented agents  & \multirow{2}{*}{(Figs. \ref{fig:AnalysisMapExamples},  and Fig. \ref{fig:Segregation})}  \\
  ... of value-oriented agents      &                                         \\
  ... of ethnicity-oriented agents from majority ethnicity  &  \multirow{4}{*}{(Fig. \ref{fig:SegregationMinority})} \\
  ... of ethnicity-oriented agents from minority ethnicity  &                       \\
  ... of value-oriented agents from majority ethnicity  &                       \\
  ... of value-oriented agents from minority ethnicity  &                       \\
  Value segregation $\Theta^\text{V}$        & mean value segregation    \\ 
  ... subgroups as above & (Figs. as above)\\
 Neighborhood density $d$                   & mean neighborhood      \\  
  ... subgroups as above as above & (only Figs. \ref{fig:AnalysisMapExamples}, \ref{fig:Segregation},  and \ref{fig:SegregationMinority})\\
 \bottomrule
\end{tabular}
\caption{Tables of static agent variables, parameters, parameter sweep values for simulation experiments, and output measures. Typewriter font is a reference to names used in NetLogo code. } \label{tab:parameters}
\end{table}

On the societal level, we measure the {\it ethnic} and {\it value segregation}, $\Theta^\text{E}$ and $\Theta^\text{V}$, as the average ethnic and value similarities in the neighborhoods of all agents. For a deeper analysis, we also measure the ethnic and value segregations of the subgroups of the ethnicity-oriented and the value-oriented agents, as well as these two groups further divided into the two ethnic groups. These averages are naturally only computed over agents whose neighborhood is not empty. We additionally measure the neighborhood density of agent $i$ as 
\begin{equation}
d_i = \frac{\#N(i)}{8}
\end{equation}
and the average {\it neighborhood density} of the society $d$. The neighborhood density can also be computed for different agent subgroups.
Finally, we also measure the fraction of happy agents. 

Tab. \ref{tab:parameters} lists all static and dynamic agent variables, parameters of the model and the output measures used in the following section. We provide the NetLogo model for reference \cite{PaolilloLorenz2018ValueSegregation}. 

We performed a parameter sweep in NetLogo's BehaviorSpace to explore the consequences of our extension on the aggregated results of the happiness of the agents, their ethnic segregation $\Theta^\text{E}$, their value segregation $\Theta^\text{V}$ and their neighborhood density $d$. Two datasets were collected to compare our extension with the original Schelling model. The first dataset, which is from Schelling's original model, also manipulates the relative group size but without value-oriented agents. The second main dataset is from our extension, including both value-oriented agents and varying size of the majority ethnic group. We provide both datasets for reference \cite{PaolilloLorenz2018ValueSegregation}. The results can also be reproduced by re-running the BehaviorSpace computation as set up in the NetLogo file. 

We initialized each simulation with an expected agents density of 0.7. The fraction of the majority ethnicity was swept over 50\%, 60\%, 70\%, 80\% and 90\%. The core parameter sweep was over the two different homophily thresholds $\theta^\text{E}$ and $\theta^\text{V}$. Both of these thresholds were scanned from 0 to 1 in increments of 0.1. The values of the parameter sweep are shown in Tab. \ref{tab:parameters}. For each point in the parameter sweep space, we ran 10 simulation runs. All output measures are averages over these 10 runs. Each simulation was run either until all agents were happy, or until 1000 time steps had passed. Output measures were saved for the last time step of each simulation run.

\section{Results}

\begin{figure}[th]
\begin{center}
\includegraphics[width=15.5cm]{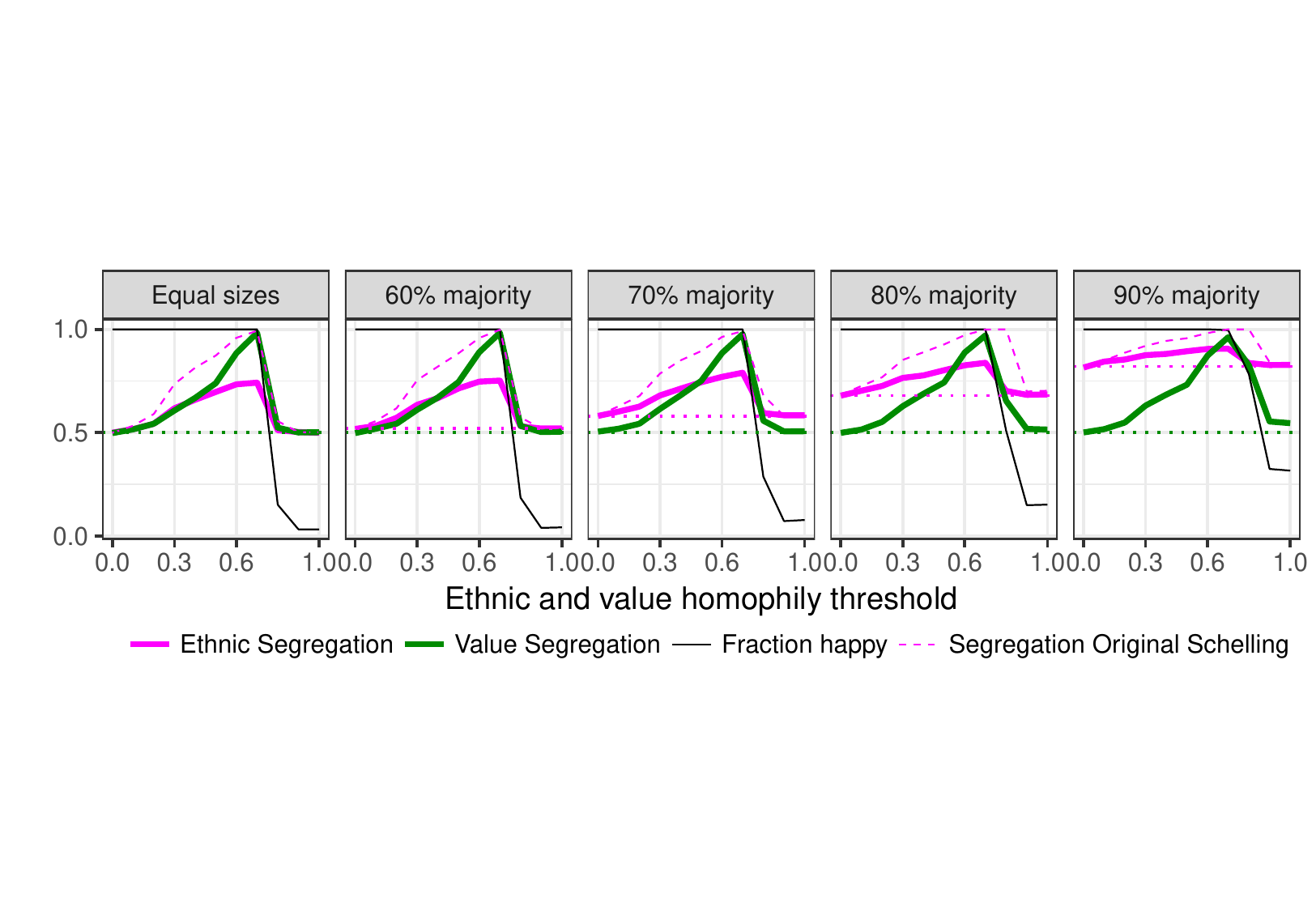}
\vspace*{-50pt}\caption{Ethnic and value segregation ($\Theta^\text{E},\Theta^\text{V}$) for equal ethnic and value homophily thresholds $\theta^\text{E} = \theta^\text{V}$ in comparison to ethnic segregation in Schelling's original model. Results for societies with different sizes of each ethnic group. Dotted lines denote the baseline segregation in random initial conditions.}
\label{fig:OverallSegregation}
\end{center}
\end{figure}

To provide a comparison, in Fig. \ref{fig:OverallSegregation} we report the emergent segregation in Schelling's original model, in which all agents are ethnicity-oriented. With an ethnic homophily threshold of $\theta^\text{E}=0.3$, an ethnic segregation of $\Theta^\text{E} =0.735$ emerges. The baseline ethnic segregation of the initial configuration is 0.5. Thus, all agents would be happy with the realization of the initial segregation in their neighborhood. This is not only the case because of random fluctuations in the initial condition. The dynamics of the repeated relocation of unhappy agents triggers the evolution of an equilibrium configuration, in which all agents are happy. The ethnic segregation of 0.735 in this configuration is larger than the initial baseline segregation of 0.5 and way beyond the homophily threshold of 0.3. 

\begin{figure}[th]
\begin{center}
\includegraphics[width=13.0cm]{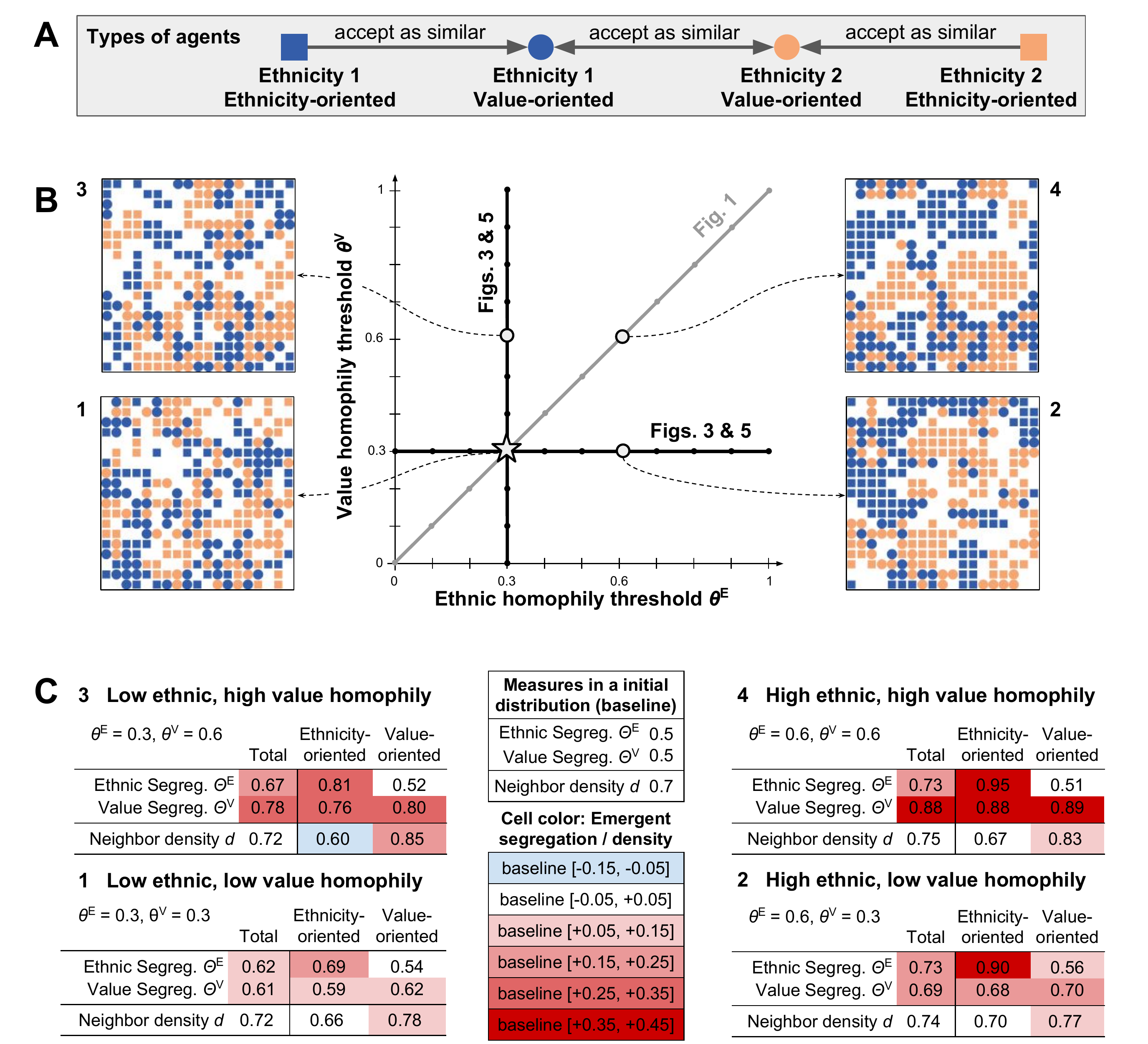}
\vspace*{8pt}
\caption{(A) Types of agents and who they consider similar. (B) The $(\theta^\text{E}, \theta^\text{V})$-space of different combinations of ethnic and value homophily thresholds. The gray diagonal refers to the parameters in Fig. 1. The black lines refer to the parameters analyzed in Figs. 3 and 5. Further, four examples of simulation outcomes for $(\theta^\text{E}, \theta^\text{V}) = (0.3,0.3), (0.3,0.6), (0.6,0.3), (0.6,0.6)$. (C) Numeric values for ethnic segregation, value segregation, and neighborhood density for the four examples in equal group size conditions. Numbers are computed once for the total population and separately for the half-groups comprising the ethnicity-oriented and the value-oriented agents. The numbers derive from the simulation analysis, whereby each number is the average over ten runs. A separation with respect to ethnicity is not of interest, since the condition that is presented is for equal group size, meaning the numbers would essentially be the same for both ethnicities.}
\label{fig:AnalysisMapExamples}
\end{center}
\end{figure}

Fig. \ref{fig:OverallSegregation} compares the emergent ethnic segregation in Schelling's original model with our extension, in which half of the agents in each ethnic group are value-oriented. In this analysis, the value-oriented agents have the same homophily threshold as the ethnicity-oriented agents, $\theta^\text{E} = \theta^\text{V}$. Further, using panels, we also show how segregation changes when one group is in a majority condition and one in a minority condition. We report the range from equal sizes to  90\% majority. The thick lines denote ethnic segregation ($\Theta^\text{E}$ in magenta) and value segregation ($\Theta^\text{V}$ in green). Examining equally sized ethnic groups and low homophily thresholds of $\theta^\text{E} = \theta^\text{V} = 0.3$, we observe that the emergent ethnic and value segregation for all agents in the population are almost equal and lower than in Schelling's original model $\Theta^\text{E} = 0.62 \approx 0.61 = \Theta^\text{V}$. To interpret the degree of ethnic and value segregation in our model, it is important to note that only half of all agents consider the ethnic or value segregation of their local neighborhood as important for their happiness. The value segregation for all agents is computed as an average over agents of which one half is ethnicity-oriented and does not care about values, and vice-versa for ethnic segregation. Nevertheless, even if we specifically look at only the ethnic segregation of ethnicity-oriented agents ($\Theta^\text{E} = 0.69$) or solely the value segregation of value-oriented agents ($\Theta^\text{V} = 0.62$), we find that both are lower than in Schelling's original model (numbers are shown in Fig. \ref{fig:AnalysisMapExamples}C). 

Higher homophily thresholds in Fig. \ref{fig:OverallSegregation} show an increase in segregation, particularly for value segregation. The sudden drop in segregation for homophily thresholds beyond $\theta^\text{E} = \theta^\text{V} = 0.7$ correlates with a drop in the fraction of happy agents from 100\% to almost zero. The reason is that, for such strong homophily preferences, the system is not able to find an equilibrium through random moves of unhappy agents. It is important to keep this order-disorder transition in mind when interpreting the following results. Fig. \ref{fig:OverallSegregation} also shows that changes in the size of the majority ethnic group leaves value segregation almost unaffected. Moreover, it influences ethnic segregation in a simple way: In a society where the majority ethnicity is a fraction of $x$ the baseline ethnic segregation in random initial conditions is $x\cdot x + (1-x)\cdot (1-x)$. (Example: 80\% of agents have an ethnic segregation of 0.8 and 20\% have 0.2.) Thus, baseline ethnic segregation under 60\%, 70\%, 80\%, and 90\% majorities is 0.52, 0.58, 0.68, and 0.82. Segregation above the baseline changes proportionally with the distance between the baseline segregation and 1.

Fig. \ref{fig:AnalysisMapExamples}B illustrates emergent equilibrium configurations. As in Fig. \ref{fig:AnalysisMapExamples}B, examples 1 and 4 show cases with equal ethnic and value homophily thresholds. For low homophily thresholds $(\theta^\text{E}, \theta^\text{V}) = (0.3,0.3)$, some clustering is visible, but still we continue to see areas in which all four types of agents are mixed. For high homophily thresholds $(\theta^\text{E}, \theta^\text{V}) = (0.6,0.6)$, areas of almost exclusively value-oriented agents are clearly visible, as well as areas of ethnicity-oriented agents for both ethnicities. Fig. \ref{fig:AnalysisMapExamples} extends the analysis to societies with a lower value-orientation and a greater ethnic-orientation and vice versa. In particular, we focus on the effects of cross-contagion in strengthening homophily preferences on increases in segregation. To that end, let us consider a society with low ethnic and value homophily preferences, as in Examples 1 from Fig. \ref{fig:AnalysisMapExamples} (see Sections B and C). When the ethnicity-oriented agents change to a high homophily threshold $\theta^\text{E} = 0.6$, as in Example 2, we observe that not only does ethnic segregation increase, as in Schelling's original model, but also value segregation $\Theta^\text{V}$ increases from 0.61 to 0.69. 
Similarily in Example 3: When the value-oriented agents increase their homophily threshold to $\theta^\text{V} = 0.6$, not only value segregation increases, but also ethnic segregation increases from 0.62 to 0.67. This scenario is entirely due to a strong increase in ethnic segregation in the group of the ethnicity-oriented agents. Thus, by increasing their value homophily preferences, the value-oriented agents have not changed their ethnic segregation, but rather increased the ethnic segregation of the ethnicity-oriented agents, with whom they do not share values. This increase of ethnic segregation in the ethnicity-oriented group is substantial, from 0.69 to 0.81. These new cross-contagion effects are the main new emergent effects in our extended Schelling model.

\begin{figure}[th]
\begin{center}
\includegraphics[width=17.3cm]{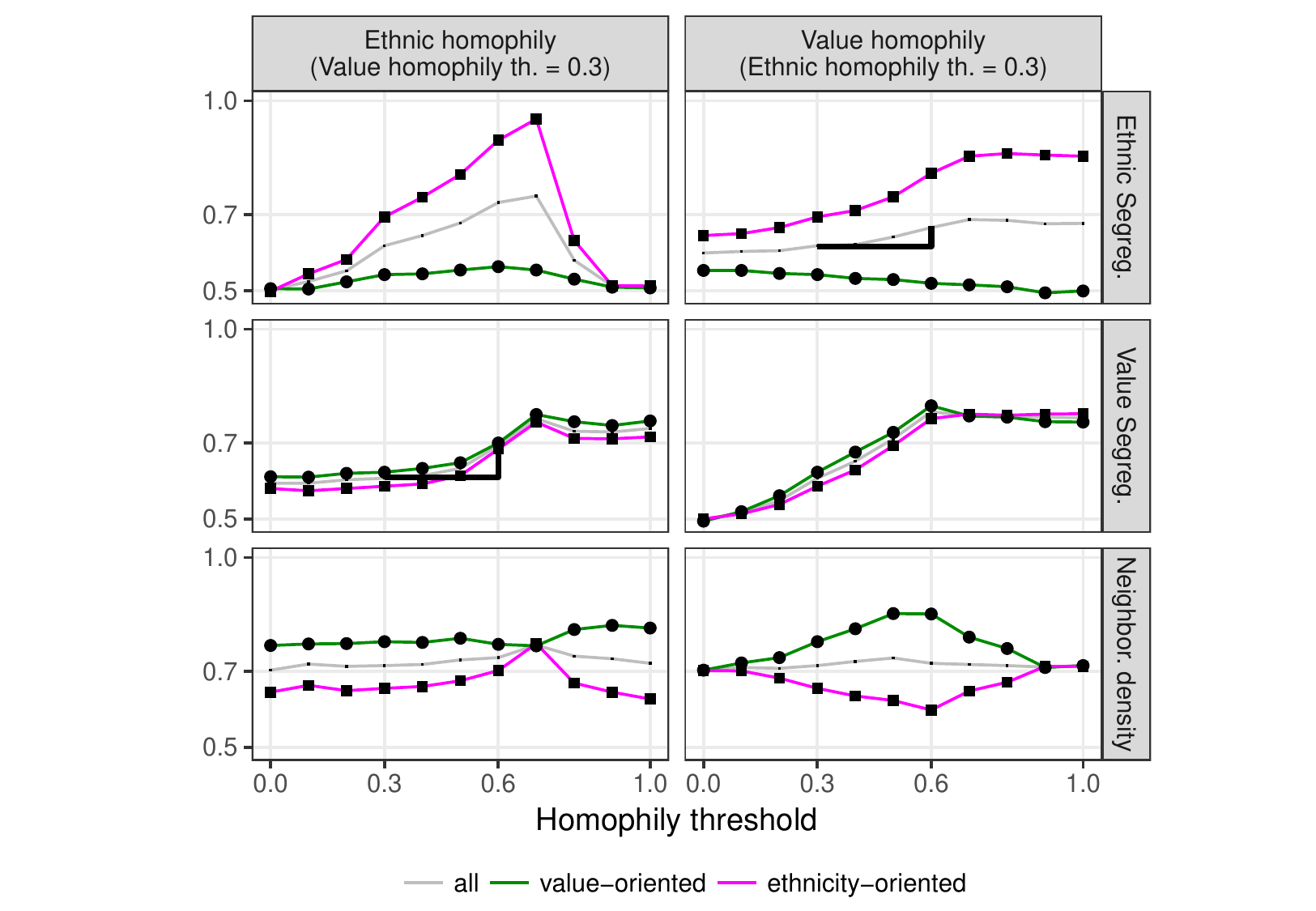}
\caption{Ethnic segregation, value segregation, and neighborhood density ($\Theta^\text{E}$, $\Theta^\text{V}$, and $d$) as a function of the ethnic and value homophily thresholds ($\theta^\text{E}$ and $\theta^\text{V}$) with $\theta^\text{V}=0.3$ and $\theta^\text{E}=0.3$ respectively (cf.~Fig.~\ref{fig:AnalysisMapExamples}) in equal sizes condition. Computation of segregation and density separated for ethnicity-oriented and value-oriented agents. In all underlying simulations both ethnicities have equal sizes. The thick black lines show effects of cross-contagion: The emergent ethnic segregation through higher value homophily thresholds and the emergent value segregation through higher ethnic homophily thresholds.}
\label{fig:Segregation}
\end{center}
\end{figure}

\begin{figure}[th]
\begin{center}
\includegraphics[width=16.0cm]{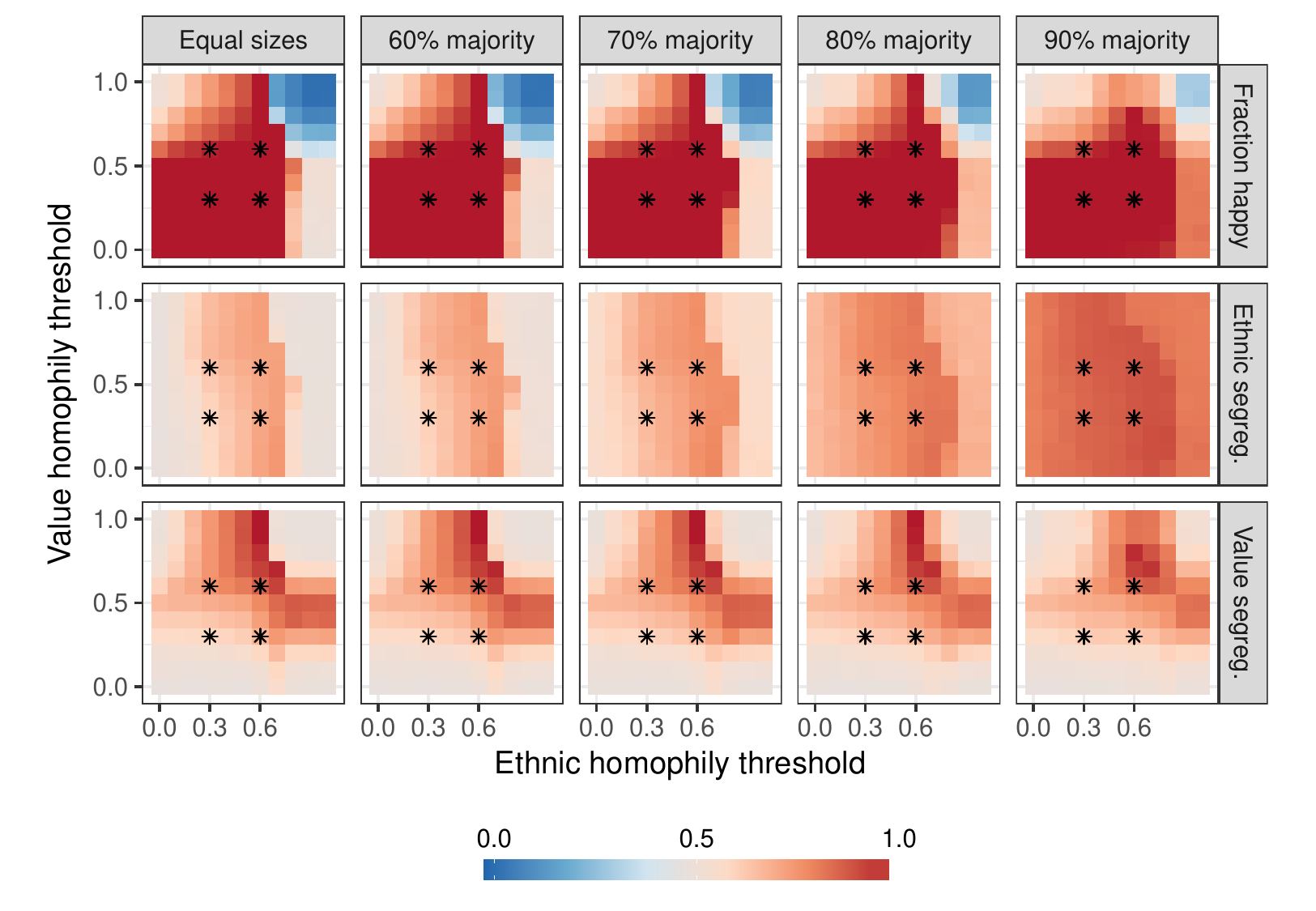}
\vspace*{8pt}
\caption{Heatmap  of the fraction of happy agents, ethnic segregation $\Theta^\text{E}$, and value segregation $\Theta^\text{V}$ with respect to the space of the ethnic and value similarity thresholds $(\theta^\text{E}, \theta^\text{V})$ (cf. Fig. \ref{fig:AnalysisMapExamples}). Panel columns show simulation results for different sizes of the two ethnic groups.}
\label{fig:Heatmap}
\end{center}
\end{figure}

Fig. \ref{fig:Segregation} demonstrates how ethnic and value segregation changes with respect to one homophily threshold when the other is held constant at a low homophily threshold, as shown by the black lines in Fig. \ref{fig:AnalysisMapExamples}B. The thick black lines in Fig. \ref{fig:Segregation} emphasize the cross-contagion effect. Further, this figure illustrates that ethnic segregation stays relatively close to the baseline level of 0.5 for the group of the value-oriented agents in all simulations. Ethnic segregation reaches very high levels for the group comprising the ethnicity-oriented agents. In contrast, the value segregation of the ethnicity-oriented agents is only slightly below the value segregation of the value-oriented agents in all simulations. Another aspect of interest is how homophily preferences affect the density of agents in their neighborhoods. In all our simulations, the baseline density in initial conditions is 0.7, meaning that, on average, 70\% of the eight neighboring patches are occupied by others. The panels at the bottom of Fig. \ref{fig:Segregation} illustrate that the total neighborhood density is only slightly above baseline. Value-oriented agents usually live with substantially more agents in their neighborhood than ethnicity-oriented agents. For a low-value homophily threshold ($\theta^\text{V}=0.3)$, this result holds for almost all ethnic homophily thresholds (see panel on the left-hand side), except for $\theta^\text{E}=0.7$, which is the highest ethnic homophily threshold for which an equilibrium of all happy agents can be reached (cf. Fig. \ref{fig:Heatmap}). For a low ethnic homophily threshold ($\theta^\text{E}=0.3)$, the difference between the neighborhood densities of the value-oriented and the ethnicity-oriented agents is the largest for intermediate value homophily thresholds (see panel on the right-hand side of Fig. \ref{fig:Segregation}).

Fig. \ref{fig:Heatmap} illustrates the fraction of finally happy agents, and ethnic and value segregation for the whole society as a heatmap in the $(\theta^\text{E},\theta^\text{V})$-plane (cf. Fig. \ref{fig:AnalysisMapExamples}B). Results for different sizes of the majority ethnicity are shown in different panels. The panels concerning the fraction of happy agents show some insights about the border of the order-disorder transition. Dark red areas mark the ordered region, in which society reaches a configuration in which all agents are happy. Interestingly, this result is possible for all value homophily thresholds when the ethnic homophily preference of the ethnicity-oriented agents is relatively high at $\theta^\text{E}=0.6$. Under these conditions, happiness for all agents can be reached even when value-oriented agents do not tolerate any ethnicity-oriented agents in their neighborhood ($\theta^\text{V}=1$). This outome is surprisingly not possible when these ethnicity-oriented agents only have low ethnic homophily preferences, e.g. $\theta^\text{E}=0.3$. In this case, some value-oriented agents remain unhappy and unable to find a place to be happy. And they do not temporarily build a configuration that makes other agents move which could create space for them to form or join a cluster of happy value-oriented agents. For the highest ethnic homophily threshold, there is no value homophily threshold for which it is possible to  reach happiness for all. With respect to the size of the majority ethnicity, the results are relatively similar with one notable further insight. For larger majorities the fraction of happy agents does not drop to zero in the disordered region ($\theta^\text{E}\geq 0.8, \theta^\text{V}\geq 0.8$). In a society with a 90\% majority, half of the agents are happy. These are mostly the ethnicity-oriented agents from the majority ethnicity (45\% of the population). These agents are easily happy because of the low fraction of agents from the other ethnicity.  For societies with ethnic groups of unequal size, it makes sense to separately examine the differences in the emergent ethnic or value segregation separately for the different ethnic groups. 

\begin{figure}[th]
\begin{center}
\includegraphics[width=28.0cm]{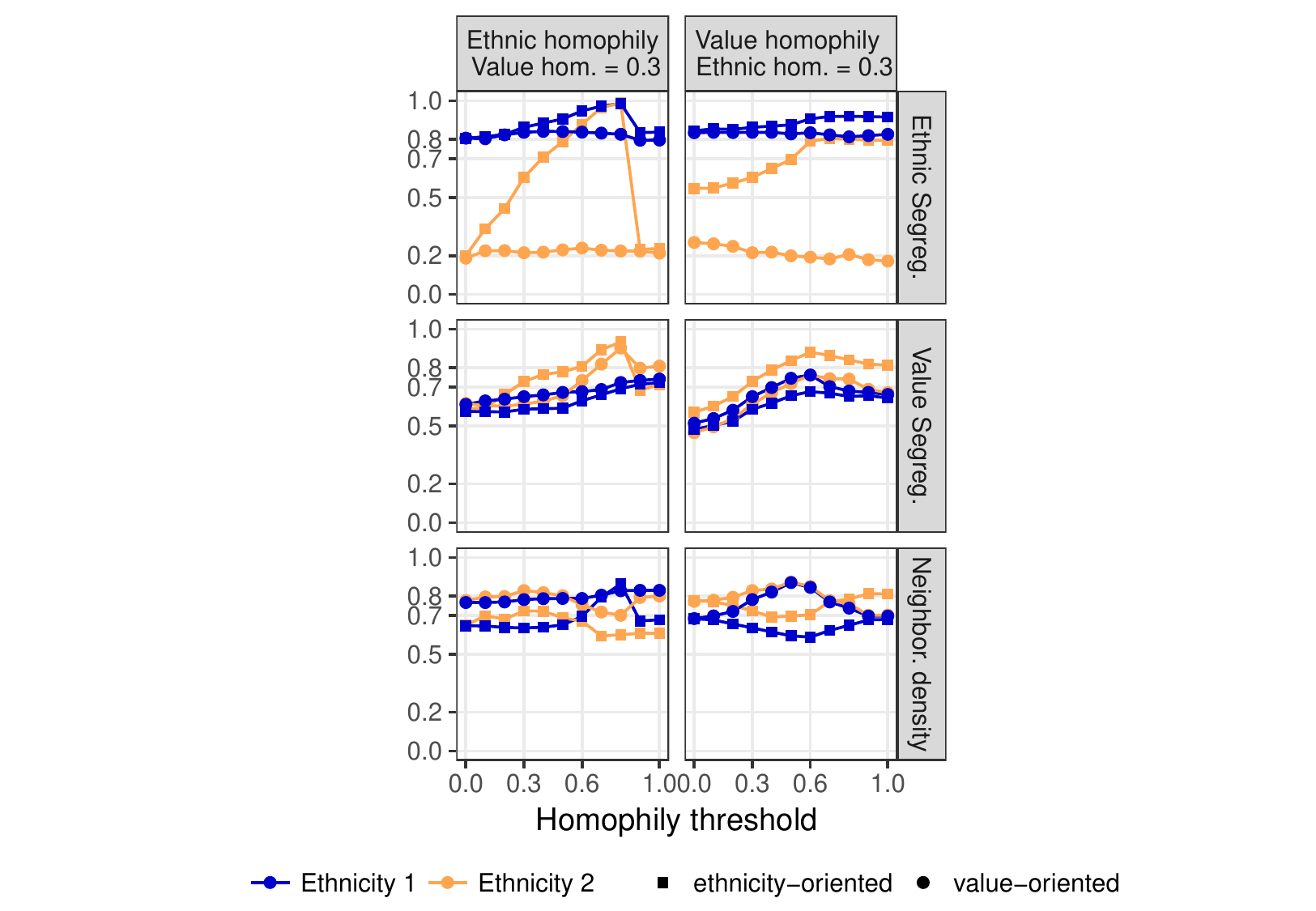}
\vspace*{8pt}
\caption{Same panel structure as Fig.\ref{fig:Segregation} but for societies with an 80\% majority of ethnicity 1 and a 20\% minority of ethnicity 2. Ethnic segregation, value segregation, and neighborhood density are computed separately for the ethnicity-oriented majority (40\% of the population), the value-oriented majority (40\%), the ethnicity-oriented minority (10\%), and the value-oriented minority (10\%). The horizontal guidelines illustrate the baseline value segregation (50\%) the baseline ethnic segregation of the minority and the majority (20\% and 80\%), and the baseline density (70\%).}
\label{fig:SegregationMinority}
\end{center}
\end{figure}

This task was not of interest in Fig. \ref{fig:Segregation} because  ethnic groups of equal size are formally identical and thus reach the same numeric values of segregation. Fig. \ref{fig:SegregationMinority} reproduces Fig. \ref{fig:Segregation} for societies with an 80\% majority of ethnicity 1 (blue). In contrast to Fig. \ref{fig:Segregation}, Fig. \ref{fig:SegregationMinority} separately illustrates the emergent ethnic and value segregations and the neighborhood densities for all four types of agents (cf. Fig. \ref{fig:AnalysisMapExamples}A) separately. To roughly summarize, increased ethnic segregation, value segregation and population density in the neighborhood appear to be more drastic for the minority ethnicity. In particular, the difference in the emerged ethnic segregation for value-oriented agents and ethnicity-oriented agents of the minority condition is striking. It is important to note that all the presented quantitative results hold only for societies with densities of 0.7 and only for the given world of $51\times 51$ patches with periodic boundary conditions (torus). Nevertheless, results would be similar for larger and slightly smaller worlds. Higher densities would probably affect the critical value of the order-disorder transition. Under higher densities it is generally more difficult for a society to reach happiness for all agents because of the limited space to move.

\section{Conclusions and Discussion}
In a population of individuals from two ethnic groups of which half are value-oriented and define similarity as shared values of ethnic tolerance, ethnic segregation is generally reduced when compared to Schelling's original model. This result  is because the value-oriented people of both ethnic groups tend to separate from ethnicity-oriented people in ethnically mixed regions, while ethnicity-oriented people tend to situate in regions divided by ethnicity. The ethnically mixed region interestingly shows a higher neighborhood density. It is possible that we have identified a driving force that could provide an explanation for the empirical phenomenon that ethnic diversity as well as ethnic tolerance is higher in densely populated areas (e.g. cities), while regions with low population density often show both less ethnic diversity and less ethnic tolerance. We do not claim that the mechanisms in our model are the only explanations for the differences in diversity and tolerance values between cities and rural areas, but we think their emergence in such a simple model is notable. 

Further on, our results suggest interesting cross-contagion effects which can shed some light on the unintended consequence of individuals strengthening their homophily preferences. 
When tolerant people increase their homophily preferences with rejection of intolerant attitudes, we observe an increase in ethnic segregation over time. This finding is remarkable because this increase in ethnic segregation is neither due to any preference changes in the people who care about ethnicity, nor is it intended by the promotion of tolerance.
When ethnicity-oriented people instead increase their homophily preferences, e.g. through an increase in racist sentiments, this creates, over time, greater value segregation for all subpopulations. This result is remarkable, because ethnicity-oriented agents would accept tolerant others of the same ethnicity in their neighborhood, but end up with fewer of these individuals. 
In general, value segregation makes perception bias about the value distribution across the whole society more likely. In particular, ethnicity-oriented people who extrapolate from their neighborhoods (with no or few tolerant people) might severely underestimate or not even believe the degree of ethnic tolerance in society. Of course, this argument  also similarly holds for tolerant people, who might underestimate the degree of ethnic homophily preferences in society. 
In the ethnic minority, cross-contagion effects are more drastic. This result demonstrates that relative group sizes, in addition to homophily preferences and spatial constraints, play a role in Schelling-type dynamics. For the minority group, aggregating value-oriented agents render ethnically mixed neighborhoods more attractive to ethnicity-oriented agents of their group, so to ethnically segregate. This result also offers an alternative to assuming the inevitable assimilation of minority groups as suggested by Esser \cite{Esser2010}.

We recognize some areas for further modeling work. First, the attribution of values could be more realistically defined as a continuous and not only a categorical state variable. This shift can facilitate comparison with real-world data, and it theoretically means to explore if such heterogeneity of agents leads to qualitatively new phenomena. Agents could similarly pursue both ethnic and value homophily with different importance. In general, a better empirical understanding of how agents form their homophily preferences is fundamental to judge the implications of our model's results for practical application.
The regular grid might be a reasonable approximation of the real world for many purposes, but human settlement shows much greater variation in local population density than can be modeled in a grid with only one resident per patch. Multiple residents per patch might be a fruitful route to more realistically model this variation \cite{GargiuloGandicaea2017EmergentDenseSuburbs}. 

Despite these and other potential extensions and robustness tests, we think that different and overlapping definitions of
similarity are useful extensions to Schelling's model as they can aid in describing complex and apparently contradicting segregation scenarios. In modern, multi-ethnic societies people segregate not only along ethnic boundaries, but many dimensions, but many dimensions, e.g from social class to cultural norms, which can increase or decrease the distance between ethnic groups \cite{wimmer09}. From an intergenerational perspective, the degree of adaptation of previous generations from minority groups along such boundaries constitutes the antecedent for the successful investment of new generations in the receiving society \cite{Esser2010}. This has clearly implications for the cohesion of societies itself. We believe that the complex system perspective and agent-based modeling can support empirical research in this field and, in particular, that the extension of value-oriented tolerant agents in Schelling's model will find future applications in this direction.

\section*{Acknowledgments}
This project has received funding from the European Union's Horizon 2020 research and innovation programme under the Marie Skodowska-Curie grant agreement No. 713639. Rocco Paolillo benefitted from the grant above. Jan Lorenz’s work benefitted from a grant from the German Research Foundation DFG "Opinion Dynamics and ollective Decisions" LO2024.

\bibliography{references}
\bibliographystyle{plain}

\end{document}